\documentclass{nime-alternate} 

\usepackage{anonymize} 		   

\usepackage[utf8]{inputenc}
\usepackage{hyperref}
\usepackage{pdfpages}

\begin{document}


\conferenceinfo{NIME'24,}{4--6 September, Utrecht, The Netherlands.}

\title{Transhuman Ansambl - Voice Beyond Language}

%
%
%
\label{key}
%

\numberofauthors{3} 
%
\author{
%
%
\alignauthor
\anonymize{Lucija Ivsic}\\
       \affaddr{\anonymize{Sensilab, Monash University}}\\
       \affaddr{\anonymize{900 Dandenong Road}}\\
       \affaddr{\anonymize{Caulfield, Victoria}}\\
       \email{\anonymize{Lucija.Ivsic@monash.edu}}
\alignauthor
\anonymize{Jon McCormack}\\
\affaddr{\anonymize{Sensilab, Monash University}}\\
       \affaddr{\anonymize{900 Dandenong Road}}\\
       \affaddr{\anonymize{Caulfield, Victoria}}\\
       \email{\anonymize{Jon.McCormack@monash.edu}}
\alignauthor
\anonymize{Vince Dziekan}\\
\affaddr{\anonymize{Sensilab, Monash University}}\\
       \affaddr{\anonymize{900 Dandenong Road}}\\
       \affaddr{\anonymize{Caulfield, Victoria}}\\
       \email{\anonymize{Vince.Dziekan@monash.edu}}
}


\maketitle

\begin{abstract}
In this paper we present the design and development of the \emph{Transhuman Ansambl}, a novel interactive singing-voice interface which senses its environment and responds to vocal input with vocalisations using human voice. Designed for live performance with a human performer and as a standalone sound installation, the \emph{ansambl} consists of sixteen bespoke virtual singers arranged in a circle. When performing live, the virtual singers listen to the human performer and respond to their singing by reading pitch, intonation and volume cues. In a standalone sound installation mode, singers use ultrasonic distance sensors to sense audience presence. Developed as part of the 1st author's practice-based PhD and artistic practice as a live performer, this work employs the \emph{singing-voice} to explore voice interactions in HCI beyond language, and innovative ways of live performing. How is technology supporting the effect of intimacy produced through voice? Does the act of surrounding the audience with responsive virtual singers challenge the traditional roles of performer-listener? To answer these questions, we draw upon the 1st author's experience with the system, and the interdisciplinary field of voice studies that consider the voice as the sound medium independent of language, capable of enacting a reciprocal connection between bodies.
\end{abstract} 



\section{Introduction}
Given the universality of the human voice, and its musical and en masse creative potential, it has long been of interest in human-computer interaction \cite{Gaver1986, Clark2019} and at NIME \cite{kleinberger_voice_2022}. This significance of voice interaction reveals itself wherever screen-based interaction is ineffective or limited. Recent advancements in technologies such as voice recognition, speech synthesis, digital signal processing (DSP), machine learning (ML) and artificial intelligence (AI), are at the core of popular tools such as Amazon Alexa, Apple's Siri, and Google Home. Yet, they all understandably use language-based voice interaction. On the other hand, although less present in the field of more experimental and expressive interactive works, even there language-based voice interaction still takes primacy \cite{Desjardins2021}. 

Rather than analysing or conceiving the voice as the mere language carrier, and therefore using it for clear commands to receive feedback, this paper looks at the alternative yet numerous poetic and expressive possibilities of the \emph{singing-voice}. More specifically, while the technological choir performs songs that contain lyrics, the focus of our investigation is directed towards the singing-voice as a medium in itself along with the spatiality of the work and how it enhances immersion in the act of listening and feeling the voice, both for the performer and the audience. By surrounding the human performer and the audience with an artificial choir, our investigation extends beyond the solely human voice, considering the ``voices'' of machines with their own autonomy, exploring contemporary ways of live performance.

Our work is influenced by recent philosophical, scientific and critical studies, particularly Voice studies that focus on the voice as a medium in itself, outlined further in the next section. Section \ref{s:background} offers a brief review and showcase of voice in interactive art and performance, accompanied by relevant examples that inspired our work. Then we present the \textit{Transhuman Ansambl,} its system, concept, technical features and modes of interactions. In Sections \ref{ss:performing} and \ref{ss:standalone} we demonstrate the work in two complementary modes -- as a \href{https://vimeo.com/907744160}{collective for live performance} and a standalone interactive sound installation (See also \href{https://vimeo.com/771883042}{Making of video}). We briefly address the cultural context and overall significance of the intangible cultural heritage preservation realm that this work implies. Finally, in Section 6, we analyse our observations and propose prospective applications.

\section{Background and Related Work}
\label{s:background}

\subsection{Voice Studies}
\label{ss:voice_studies}
The continuous development of sophisticated AI and ML systems, along with their ever-increasing presence in all aspects of our lives, urges us to reconsider the relationships we develop with non-human entities. The affect of performativity carried through the contemporary human voice now extends beyond the sole exclusivity of the body. This concept of performativity, a term that initially arose within J.L. Austin’s language philosophy, was powerfully introduced into feminist theory e.g~\cite{butler_critically_1993} and applied to help think about each of those aspects as ongoing processes, rather than fixed and non-performative representations. Performativity is essential to a voice that evokes and moves from and between bodies and things -- it is this performative aspect of voice that brings affect to the foreground and establishes emotions, connectivity, and attachments between individuals and things. Hence, understanding voice and its performance in technology and the arts within a posthumanist context that de-centres humans is driving the rapidly developing interdisciplinary field of \emph{voice studies} \cite{thomaidis_voice_2015}. 

In this section, we outline a small but representative number of projects selected primarily due to their approach to voice and their direct impact on our work. While these works differ from each other in concepts and ways of working with voice, their common denominator is questioning the material qualities of voice through artistic exploration that goes well beyond words. Rather than analysing or conceiving the voice as the mere language carrier, and therefore using it to receive feedback and introduce interactivity, these works utilise the sculptural and spatial possibilities of the voice. The works that follow are not limited to their way of presenting themselves to the audience  -- the list includes interactive musical interfaces, installations, and performances -- the only limitation, if it can be called as such, is their emphasis on the singing-voice that invokes emotions.

One of the first works that drew our attention due to its explicit and reciprocal relation to voice is Alvin Lucier's renowned work \emph{I Am Sitting in a Room} (1970). Lucier recorded his own voice in a reverberant room, played this original recording in the same room to achieve another recording with the same resonances, repeating the procedure until the original voice was covered by the resonance and reverberation produced by the room. Although seemingly not resembling our \emph{Transhuman Ansambl} which is among other things performed live, what they have in common is treating the voice as a pure sound material. More specifically, while both works contain text/lyrics, when played back in a spatial environment, the semantic value of the language either loses its shape entirely, turning into a textural soundscape as in the case of Lucier's work, or falls into the background as is the case with our work. 

When spatialised and expanded into a surround environment in the form of installations, the voice (as a sound medium) starts to interplay with sculpture \cite{louvel_sculpted_2019}. 
\emph{The Forty Part Motet} is one of the acclaimed examples of a sound installation, consisting of forty loudspeakers playing back a recording of Spem in Alium (1573) by Thomas Tallis. While going through available audio-video documentation of the work, we have noticed how despite being generated in simple loudspeakers, voices used in Cardiff's work retained the ability to affect viewers and evoke emotions. 

Further into the development of our work, attempting to answer questions regarding the interface between the human performer and the virtual singers, we narrowed the search by looking only at the works that use a microphone as an interface. One such pioneering work is the \emph{Singing Tree}, created by a team of researchers at MIT in the late 90s. Designed both for a personal interactive experience and as part of a larger project titled ``Brain Opera'', the participant interacts with the \emph{Singing Tree} by singing into a microphone \cite{oliver_singing_1997}. During singing, the system analyses the voice for parameters such as the pitch and rewards the participant with audio and visual feedback if they have maintained a steady pitch. The more recent musical interface developed by Stefano Fasciani and Lonce Wyse uses vocal gestures as an alternative to traditional, physical controllers. Their voice interface mapped the dynamic aspects of vocal sound to the synthesizer's parameters \cite{fasciani_voice_2012}.


\section{Transhuman Ansambl}
\label{s:transhuman}

\begin{quote}
    ``...everyone is a listener of others and a performer to others.''
    \flushright --- Atau~Tanaka, quoted in \cite{tanaka_sensor-based_2011}
\end{quote}

\begin{figure*}[h]
  \centering
  \includegraphics[width=\linewidth]{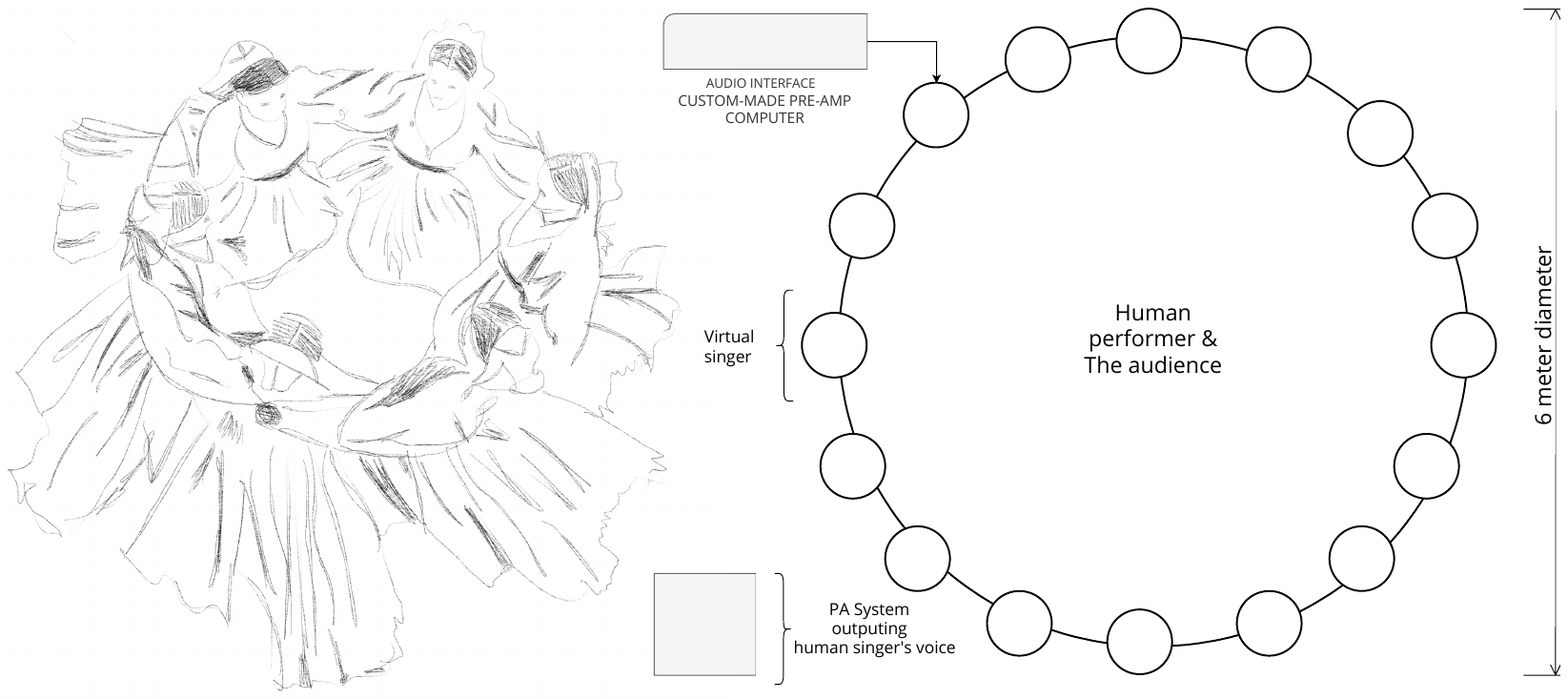}
  \caption{Installation layout sketch (left) showing the inspiration for the circular shape of Transhuman Ansmbl from traditional Croatian singing. The diagram on the right shows the layout and key elements used in the work.}
  \label{fig:circular-shape}
\end{figure*}

\emph{Transhuman Ansambl} is an interactive technological choir consisting of 16 non-human autonomous entities called \emph{virtual singers} that sense their environment and respond to a human artist’s voice and audience presence using vocalisations (explained in detail in Section \ref{ss:modes_of_interaction}). Each one of these 16 presence-sensitive singers is an individual, self-contained agent: a small and delicate object with an embedded loudspeaker, ultrasonic distance sensor, LED ring, and custom electronics developed specifically for the project. Inspired by the traditional Croatian ring-shaped social formations such as \emph{kolo}, which are defined by a strong sense of shared identity, and \emph{bećarac}, an emergent minimal vocalic chant performed as an interaction between an individual and the group, the virtual singers are circularly distributed around the performance space (Fig.~\ref{fig:circular-shape}). Designed for both live performance and as a standalone sound installation, the audience is always invited to experience the work from inside the ring, surrounded by the singers  (Fig.~\ref{fig:transhumanansambl}).

Conceptually, the work re-imagines the first author's work as a musician, lead vocalist and performer through technology, while exploring the multiplicity of her multicultural identity (a Croatian woman now living in Australia). Given her extensive practice as a singer, the work is centered around the singing-voice -- considering both the material qualities embedded in the physical voice, so much as its role in the identity-making process. Stemming from the advent of migration from Croatia to Australia, traditional Croatian songs were used as starting points for creative departure and development. Although these initial musical pieces we have created with the virtual singers borrow vocal techniques found in Croatian folklore, such as the two-part singing and call-and-response method \cite{milacic_sveuciliste_2021}, they are original compositions rather than a prescribed performance of a renowned traditional piece. Rather than introducing additional disjuncture from tradition, these hybrid musical compositions allow the first author to create space for artistic subjectivity and potentially new aesthetics.

\begin{figure}[h]
  \centering
  \includegraphics[width=\linewidth]{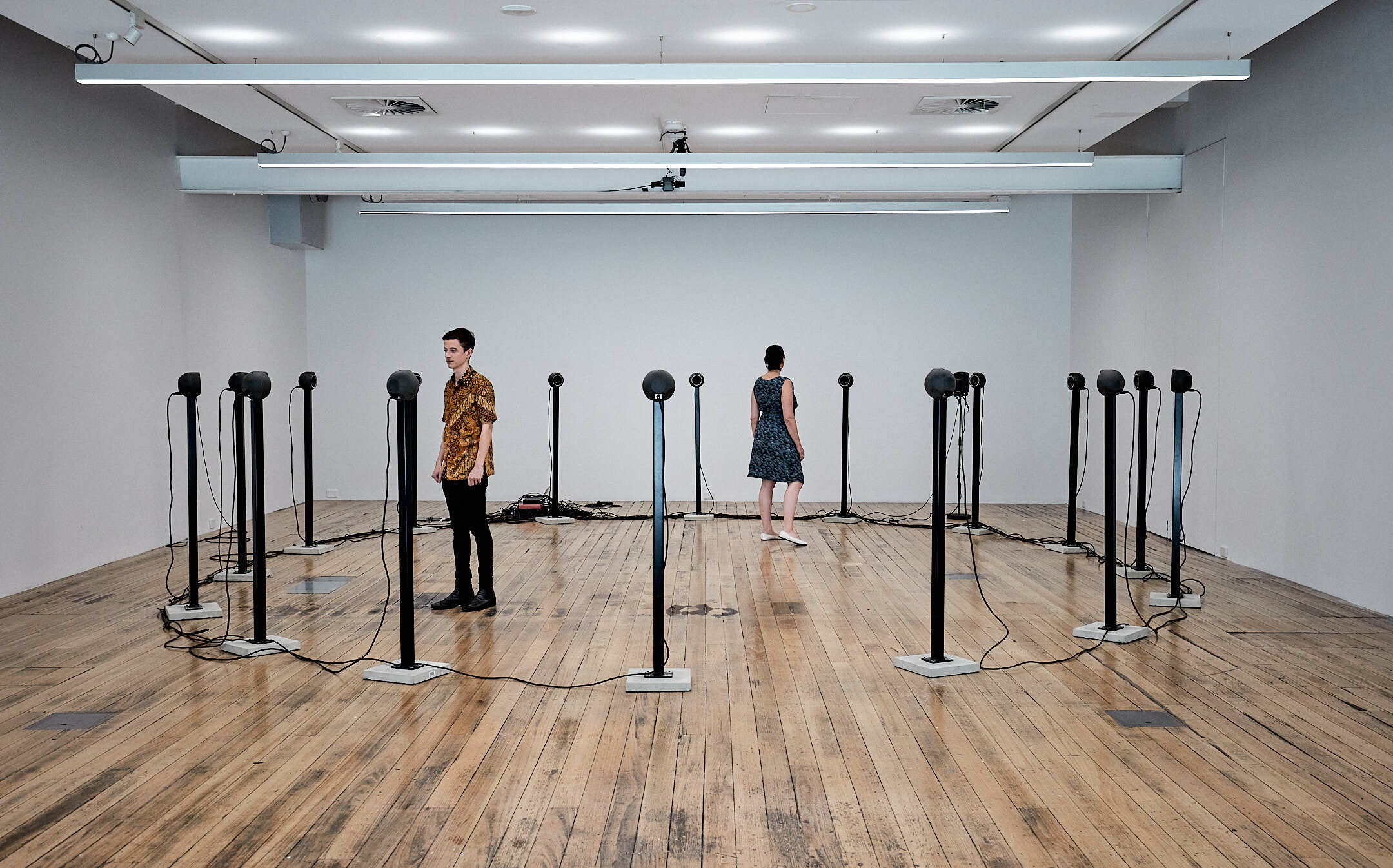}
  \caption{Standalone installation version of Transhuman Ansambl, with people interacting with the virtual singers. Melbourne, January 2024. }
  \label{fig:gallery_ansambl}
\end{figure}

\subsection{Cultural context and Ethnographic Study}
\label{s:cultural_context}

To obtain a more genuine understanding and richer knowledge about \emph{kolo} and \emph{bećarac}, throughout June 2023, the 1st author carried out ethnographic research that focused on observing two official associations dedicated to the preservation and performance of traditional Croatian folklore. During the ethnographic study, several active performers and both art directors (also acting leaders of the ensembles and once performers) were interviewed individually. The data acquired from these interviews, as well as the accompanying audio-visual observations of the groups during their rehearsals mostly consisted of information about the group's dynamic, interactions before and during performance, and level of individual autonomy.

Conversations with the artistic directors revolved around the processes of defining and appointing roles to each of the performers within the group, degrees of their autonomy, and definitions of the \emph{voice} within the group - both on an individual and group level. What are the main characteristics of a successful traditional folklore performance group? Is there a feature such as a group's lead singer, and if so, how is it defined and appointed?  

On the other hand, questions directed to individual performers focused on ways they interact with the audience, among themselves, and also if there is any room for individuality.  
What was apparent as a shared element, and as such considered a key observation extracted from both the performers' and artistic directors' answers is the highly desired uniformity, in both the aural and visual aspects of the performance. 

Expanding on previous research from the field of Intangible Cultural Heritage (ICH), structural elements such as the layout and order of the performers, as well as the reasons to sometimes form smaller, internal formations (duets or triplets), were then used as a basis to constitute system stories for the \emph{virtual singers} in the \emph{ansambl}, as explained in the next section.

Consisting of the immaterial expressions of one's culture, ICH serves to help constitute the cultural identity of its creators \cite{10.1093/ejil/chr006}. Continuously living and evolving, this heritage is based on collective memory and is a powerful device for telling stories and reaffirming one's identity \cite{goulding_heritage_2009}. Nowadays, new technologies represent a vehicle for the safeguarding and transmission of ICH and as such are yet to be fully explored and implemented through practice \cite{alivizatou-barakou_intangible_2017}.

\subsection{Virtual Singers}
\label{ss:virtual-singer}

\subsubsection{Conceptual definition}
\begin{figure}[h]
  \centering
  \includegraphics[width=\linewidth]{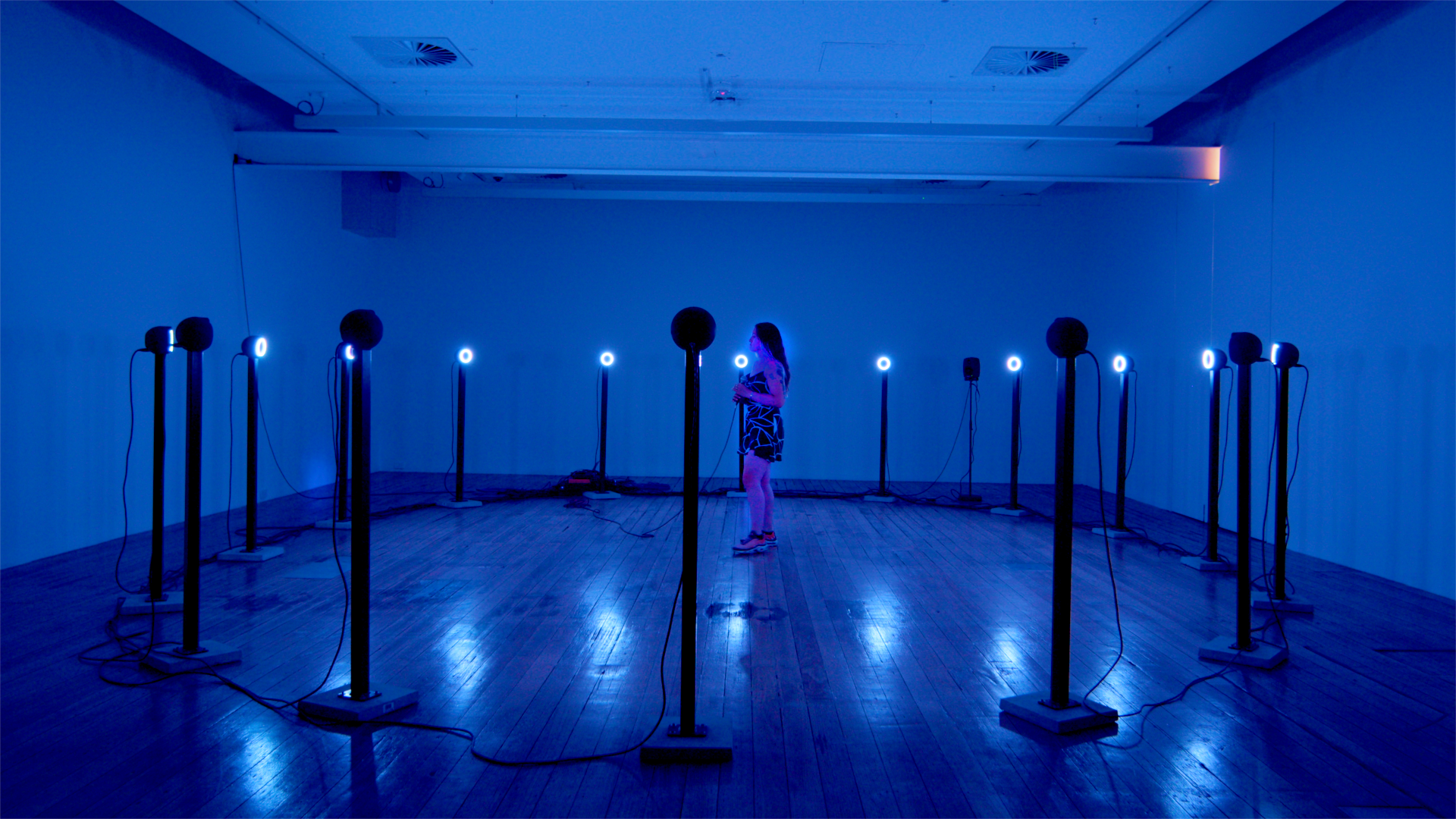}
  \caption{Excerpt from the live performance showing the uniformity in the hardware design of the virtual singers along with the position of the human performer.}
  \label{fig:allthesame}
\end{figure}

Acting as key entities in the artificial life of the \emph{ansambl}, ones that \emph{react} to their environment, the audience, and the performer, virtual singers are considered to be \emph{agents}. The notion of agency can be defined through the mere technical lens as the capability of perceiving the environment through sensors \cite{russell_artificial_1995}, and responding in a timely fashion to changes that transpire -- something virtual singers are capable of doing. 
To be more precise, in most artificial life systems, what constitutes an agent are two important features: i) level of autonomy which implies the use of sensors and constitutes the relation between perception and response, and ii) level of adaption to changes in the surrounding environment \cite{Beyls2007InteractionAS}. For virtual singers, this structure is apparent through the following features:

\begin{itemize}
    \item{\textit{Stereo pairs.} Traditionally, within the Croatian folklore group, each performer has their own pair with whom they form a duet. This was translated and constituted in our work through the creation of stereo pairs, giving each virtual singer a corresponding counterpart;}

    \item{\textit{Audio-visual uniformity.} As observed throughout the ethnographic research mentioned in the previous section \ref{s:cultural_context}, one of the key elements of a successful folklore group is uniformity on stage during live performance. This uniformity is seen as a sign of discipline and collectiveness, transpiring to the audience both through the visual content (dress, body language, height, movement) as well as the aural (singing). Hence, the hardware design of virtual singers is uniform and the same across all 16 singers (Fig.~\ref{fig:allthesame});}

    \item{\textit{Limited autonomy.} When performing traditional choreographies, each performer has a strictly dedicated role which requires rigorous adherence. Yet when not performing as a collective, individual expressiveness that varies from performer to performer is welcomed. When applied to our \emph{ansambl} and its standalone installation mode, each of the virtual singers has some minute internal decision-making mechanism that determines what and when a sound will be produced.}
    
\end{itemize}
Therefore, virtual singers have the following characteristic properties: while they are all identical at their core, each one of them belongs to a set of pre-defined types that conditions certain behaviours. These inherent ontologies then present specific attributes -- how each singer relates to the others, their relation to the human performer, their likeliness to interact, and their level of activity.

\subsubsection{Technical Details}
\label{ss:technical_implementation}
\begin{figure}[h]
  \centering
  \includegraphics[width=0.5\linewidth]{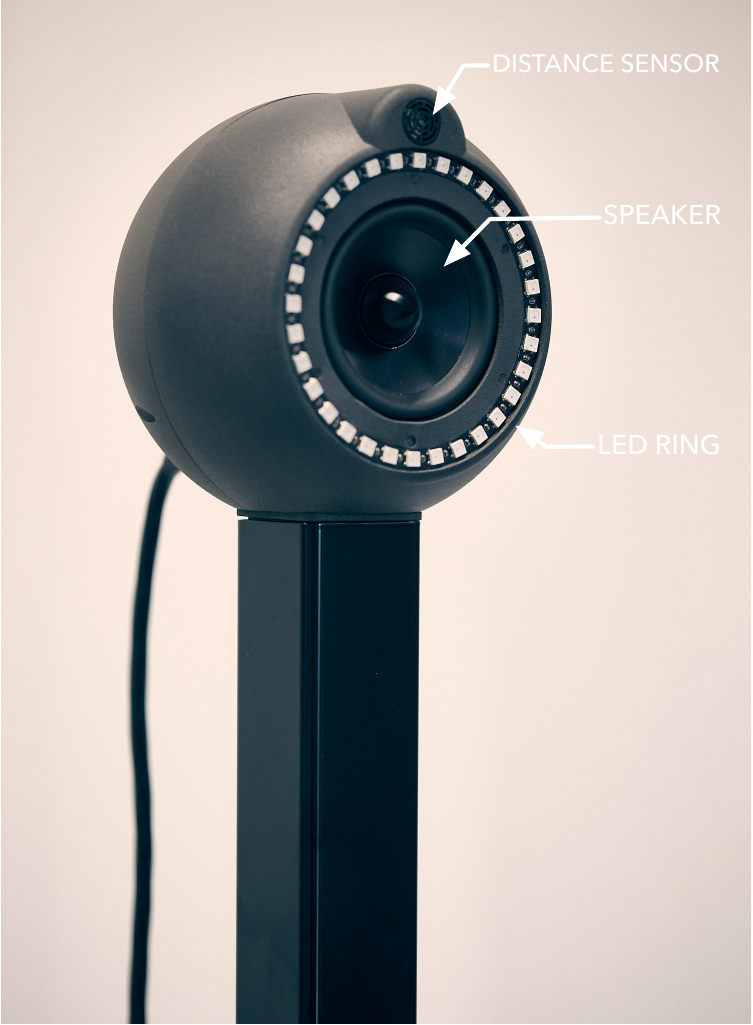}
  \caption{Closeup image of the individual singer, showing the location of the speaker, ultrasonic distance sensor, and LED light ring.}
  \label{fig:singer-detail}
\end{figure}

 Each individual singer is equipped with a speaker, a custom-made RGBW LED light ring, and an ultrasonic distance sensor (HRLV-MaxSonar-EZ0), see Fig.~\ref{fig:singer-detail}. Singers are cabled back to a central amplifier and an Arduino MEGA microcontroller. The microcontroller receives data from the distance sensors and controls the LED lights around each speaker (individual LEDs within each speaker are addressable). A central computer manages sound distribution to each speaker and controls the Arduino. A vocal microphone is used by the performer to interact with the system. The physical form of the singers was fabricated using 3D printing and the virtual singers were hand assembled in our lab before being deployed for live performance.

A singing-voice dataset was developed as part of the core software framework for the work. The dataset consists of over a hundred recorded samples of the artist's voice and was used to build a ``vocal matrix'' of sounds that are triggered by the system in response to both performer vocalisations and audience behaviour. Across the vocal dataset, three main vocal techniques were used, where each one was performed twice, in first and second voice:

\begin{itemize}
    \item{\textit{Falsetto.} A vocal technique is used for hitting higher notes than what the singer is commonly able to achieve. Results in a voice that sounds airy and soft;}

    \item{\textit{Belting.} Defined by a strong but warm voice where the singer combines their chest and mixed voice;}

    \item{\textit{Musical phrasing.} The vocal technique allows the vocalist to create a sequence of notes that would allow phrase expression (e.g., lyrics, spoken word, vocalisations).}
\end{itemize}
The dataset was developed based on traditional methods of Croatian composition, using two-part singing \cite{milacic_sveuciliste_2021}.
This dataset was then analysed and categorised manually according to pitch, vocal technique, length, and volume, then distributed among the singers, dividing them into two equal groups of eight, where one group sings the first voice, and the other one the second voice.

\subsubsection{Modes of Interaction}
\label{ss:modes_of_interaction}

As mentioned at the very beginning of Section \ref{s:transhuman}, the \emph{ansambl} exists in two modalities -- as a live performance with a human performer, and as a standalone sound installation. The installation layout and modes of interaction were inspired by research on the human voice beyond language, considering voice as an extension of the body: something that has both a physical and spatial dimension \cite{dolar_voice_2006, neumark_voice_2010}.

The work was developed in MAX MSP, visual programming software for music and multimedia. Each \emph{singer} consists of four key and connected objects that determine their state and the current mode of interaction. Those objects are the same and standardised across all 16 singers and are as follows:

\begin{itemize}
    \item{\emph{State indicator}. Based on a simple toggle between active (1)/inactive (0) states, this object determines whether the singer is singing. This state is dependent on the artist's vocal pitch and incoming ultrasonic distance sensor data. }
    
    \item{\emph{Playlist}. List of individual vocal samples created by the artist. They are assigned to each singer according to the voice group they belong to (first or second voice).} 

    \item{\emph{Ultrasonic distance data list}. Object used to register data collected from the ultrasonic distance sensor in real-time. This data is registered in the form of numbers from 1 - 10, representing the proximity of the closest object to the singer's speaker. This element was introduced as a way to give virtual singers agency apparent through the ability to ``sense'' the audience at a close distance.}

    \item{\emph{Serial RGB LED ring lights controller}. Object controlling LED ring lights, whose activity corresponds to the singer's \emph{State indicator} object. This feature was added during later stages of development to help detect which virtual singer is active (when acting individually), but also to amplify the engaging effect the circle has on the audience when live performing.}
\end{itemize}

\begin{figure*}[h]
  \centering
  \includegraphics[width=\linewidth]{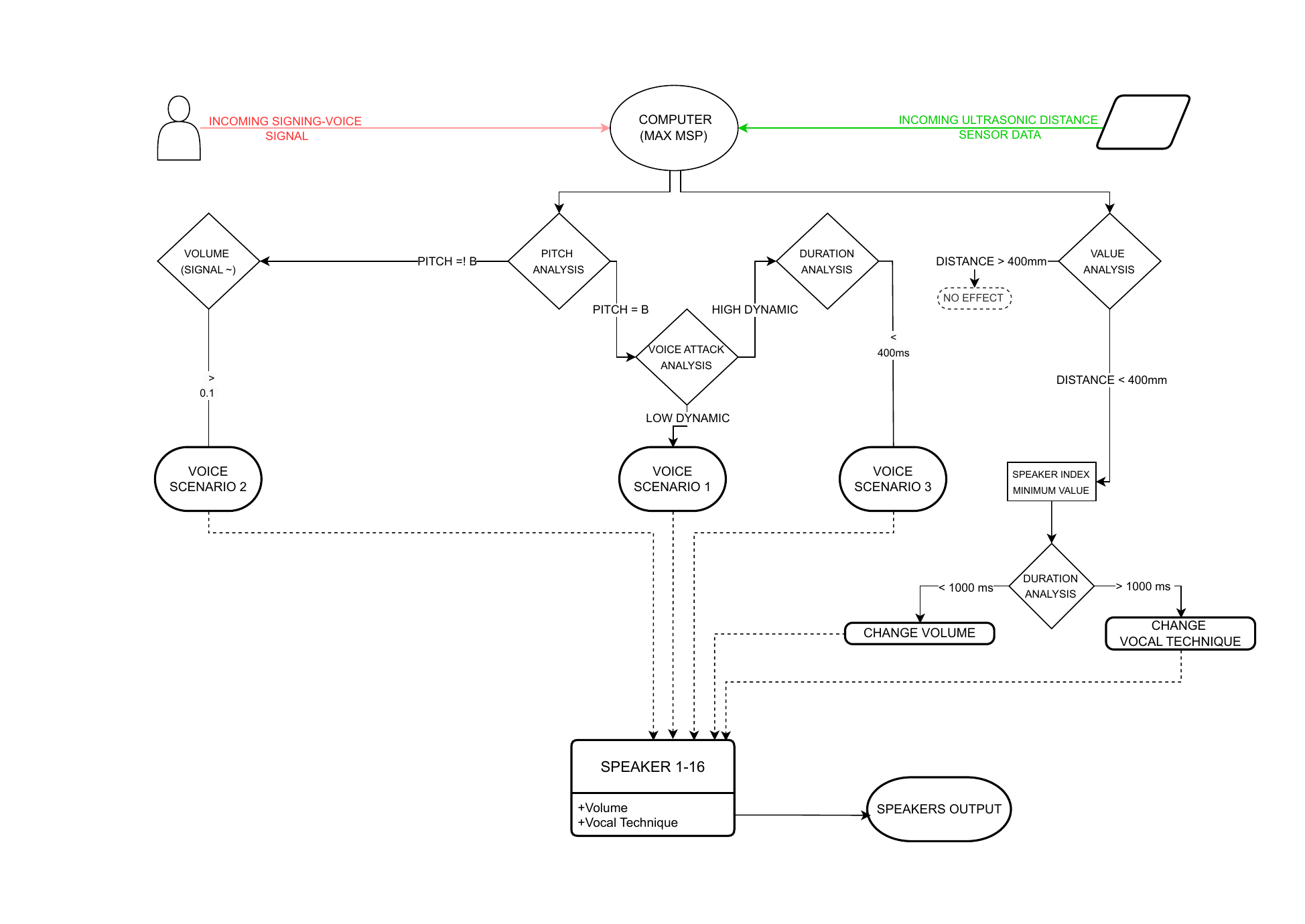}
  \caption{Transhuman Ansambl's simplified schematic diagram illustrating the interaction process and the interrelationship between the singer's vocal input and variables from the ultrasonic distance sensor on the speakers final output.}
  \label{fig:interaction-schematic}
\end{figure*}

As shown in Fig.~\ref{fig:interaction-schematic}, the artist's vocal audio is sent to the computer via a microphone, where it is analysed in real-time. This analysis includes the registration and interpretation of vocal features such as volume, pitch, and voice attack which is measured as the range of energy heard in an accordion frequency band (i.e., high dynamic results are classified as short or strong attacks while those with lower dynamics are classified as long and soft). In parallel, data from the ultrasonic distance sensor is registered and sent to the computer, affecting the output of the individual speaker by changing its volume or chosen vocal technique.
To avoid the activation of the system while the singer is silent, or speaking instead of singing, certain spectral frequency limits were set by sampling the artist's voice in various situations during development and noting its corresponding spectral signature. 
Once we have determined the spectral frequencies within which singing occurs, the software can recognise the artist's voice when singing and differentiate singing from speaking. The received datasets are then registered and analysed for aforementioned parameters; if their values are within the predetermined margins, corresponding scenario activates and accordingly sends out information to all 16 singers (Figure \ref{fig:interaction-schematic}). Each singer then responds accordingly, based on singing pitch and pitch length, selecting a vocal sample to play based on this information. This creates a type of ``call and response'' interaction between the performer and singers who respond to the performer's voice with their own voices. 

\subsection{Performing with the Ansambl}
\label{ss:performing}

\begin{figure}[h]
  \centering
  \includegraphics[width=\linewidth]{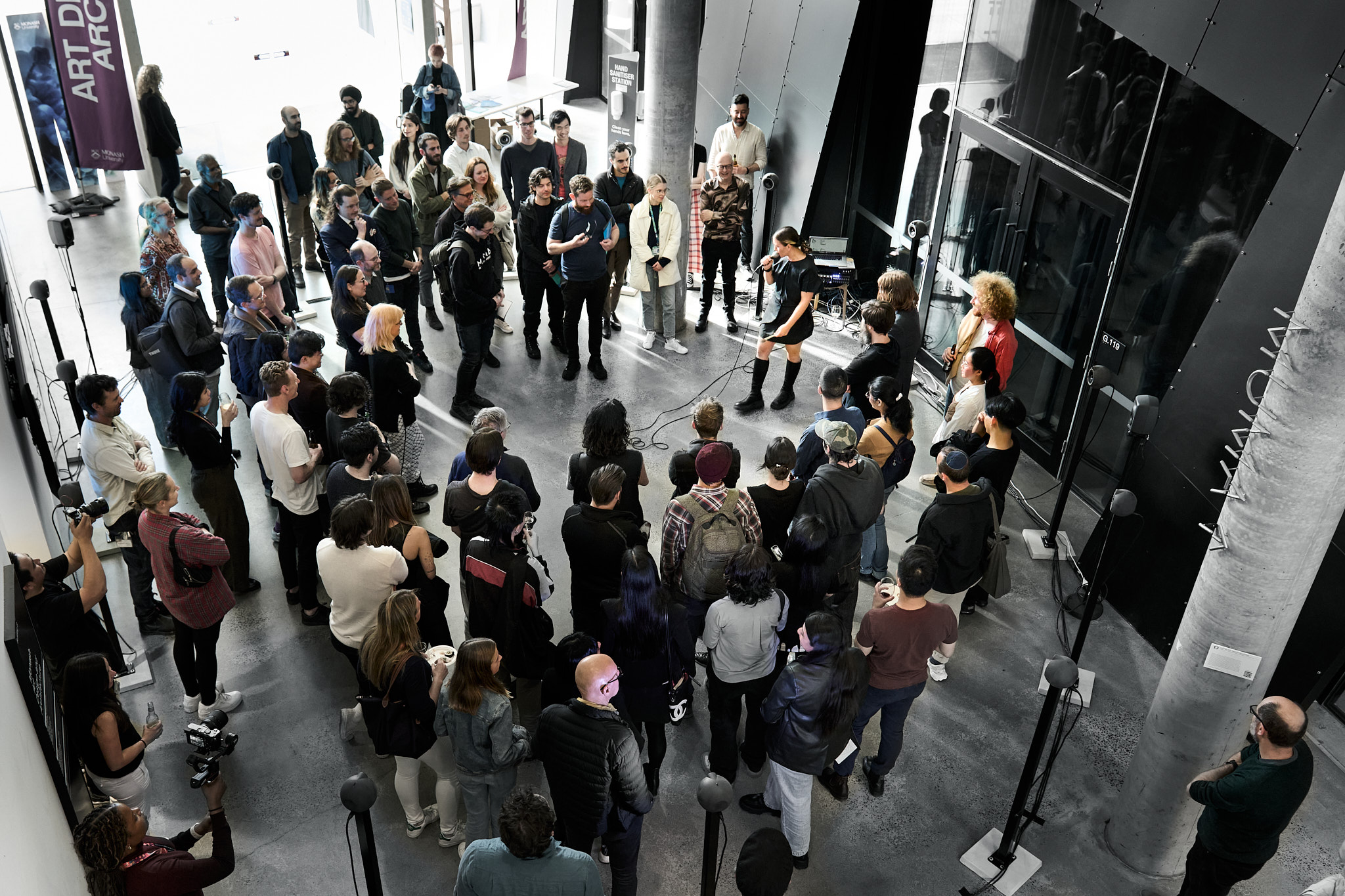}
  \caption{Transhuman Ansambl installation and live performance, Melbourne, November 2022. The audience and performer are surrounded by a circle of 16 autonomous virtual singers who collaborate and interact with the performer by listening and responding to the performer's voice. }
  \label{fig:transhumanansambl}
\end{figure}

Transhuman Ansambl was exhibited and tested through a series of performances in late November of 2022 while still in an early prototype stage Fig.~\ref{fig:transhumanansambl}. The audience (of approximately 100 people) expressed curiosity and wonder towards the singers as they were initially perceived as common speakers until becoming active. Additionally, the invitation to experience the live performance within the boundaries of the circular ring (i.e.~being surrounded by the singers) surprised the audience and evoked more interest in ``who'' the singers are and what they ``do''. Audience members noticed how their physical distance from individual singers affected the choir's singing. This was noted as an important aspect of interaction that needs further exploration since it demonstrated a potentially new way of interaction between the audience, singers and the performer. At that point, the audience became active performers, having the ability to make changes in singers' performance based on their distance from them. For example, approaching a singer resulted in changing their voice from signing to falsetto to whispering.

Consequently, the act of ``enclosing'' the audience within the boundaries of the ring in close proximity to the performer created an intimate space that connects the audience and performer. This was evident through many oral accounts expressed immediately after performances by both the artist and the audience. As mentioned in Section \ref{ss:voice_studies}, by doing so, the intersubjective and affective quality of voice was brought to the foreground.
Transhuman Ansambl suggests a newly formed human-machine assemblage that employs technology to challenge the future of traditional roles, such as that of performer-listener \cite{heim_audience_2015}, and to make us more perceptive to our surroundings, and of ways we affect each other through sound and physical presence.

The second rendition of the live performance mode is currently in the finishing stages of development and although not yet exhibited to the public, it has been tested separately with three experienced singers and the formal case study of the system is underway. 

Apart from the substantial expansion of the singing-voice dataset assigned to each of the virtual singers, the greatest addition is the option for the human performer to create an infinite, stackable feedback loop together with all 16 virtual singers. The human performer is now able to instantly record and playback their singing through each of the speakers. While live looping is a well established and popular technique, our implementation is innovative due to the following features:

\begin{itemize}
    \item{\emph{Partial control}. While the human performer is in control of what they will sing, each of the virtual singers is deciding (within a limited range) which part of the singing will be recorded and then playback through that particular audio channel. In essence, this loss of control over each ``singing loop'' creates space for surprise and improvisation, yet provides the virtual singers with an additional sense of agency and autonomy. }
    
    \item{\emph{Spatial dimension}. Human performers can (re)arrange the ongoing feedback loop simply by standing closer to the virtual singers whose playback they prefer. By doing so, the other 15 virtual singers will echo the ``chosen'' one and perform as a collective. Ultimately this allows the human performer to then use the \emph{ansambl} as an accompanying choir.} 

    \item{\emph{Immersiveness}. The human performer is not required to use any screen-based interface or device other than a microphone to perform with the virtual singers. Additionally, the spatiality of the work apparent through the 16 virtual singers that surround the human performer, reinforces their presence and as such strengthens the immersion.} 
\end{itemize}

\subsection{Standalone Sound Installation}
\label{ss:standalone}
\begin{figure}[h]
  \centering
  \includegraphics[width=1\linewidth]{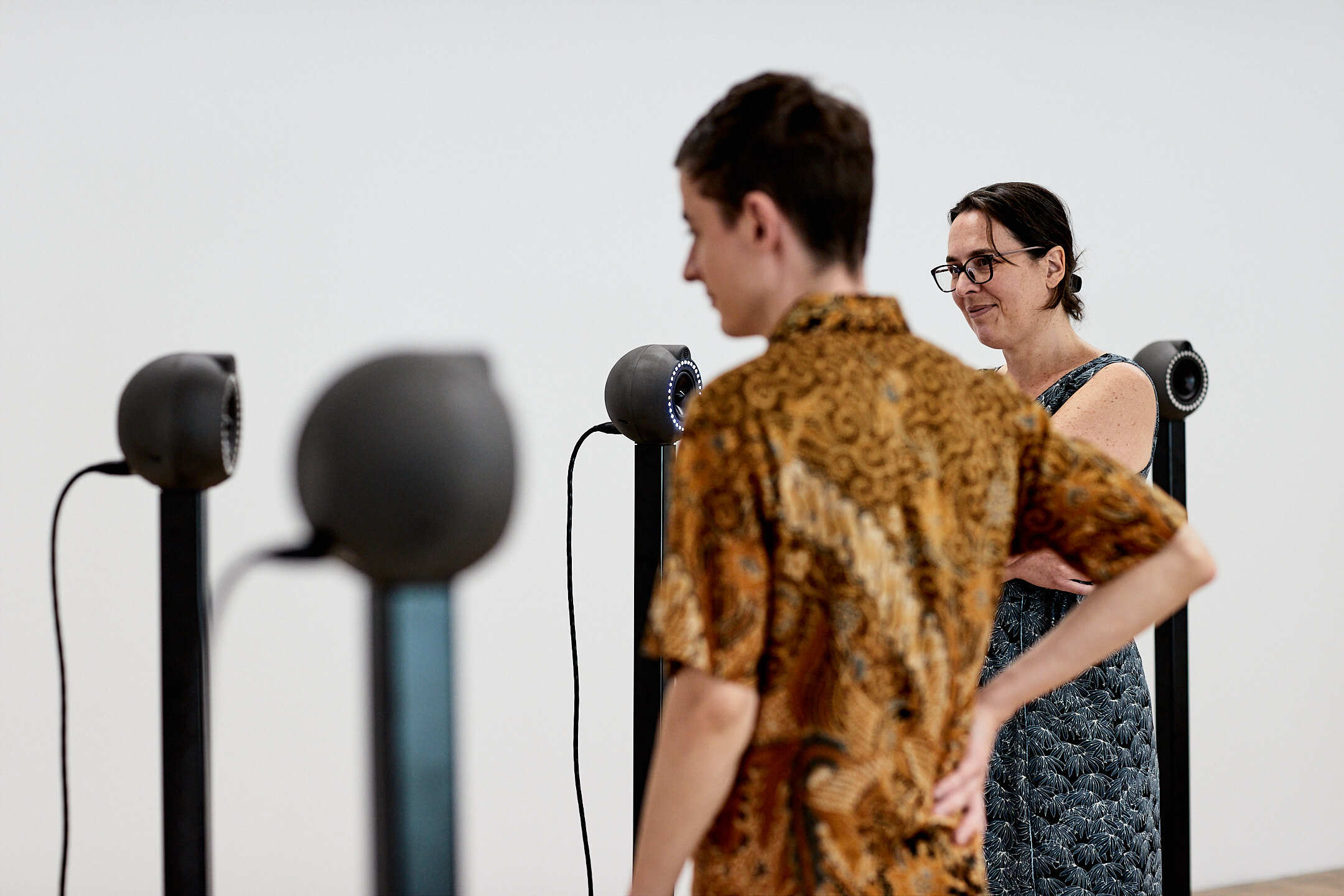}
  \caption{People interacting with the virtual singers.}
  \label{fig:interaction}
\end{figure}

The \emph{ansambl's} exhibition mode as a standalone sound installation (i.e. whenever not performing live with a human performer) is in its final stage of development. Conceptually, this rendition was envisioned as a technological translation of an intermission period in between performances among a group of performers. Moments when performers relax, engage in small talk, or prepare for the performance, which are usually hidden from the public are now brought closer to the audience. As such, it allows the audience to get closer to each of the virtual singers (Fig.~ \ref{fig:interaction}), and freely explore them. The human interaction with the singers is made possible through either mere movement within the circle, or by closely inspecting one of them, listening to the voices they produce, individually or jointly. Technically, this is where the ultrasonic distance sensor comes into play, sensing the viewer's movement and proximity. At moments the viewer can directly affect and provoke a particular singer by approaching it closely, but this won't necessarily be a rule. As explained in Section \ref{ss:virtual-singer}, virtual singers have a certain degree of autonomy which allows them to decide to respond or not.   

Hence, the modes of interaction between the audience and the virtual singers are intentionally not obvious at the beginning of the encounter; the singers produce sound but inconsistently, aiming to provoke exploration and capture the viewer's attention. 
 
During the aforementioned ethnography study in Section \ref{s:cultural_context}, we observed the behaviour of the performers during those short periods of break and noticed a variety of common actions. Breathing exercises, vocal warm-ups, casual chatter, laughter, hydration, quick phone calls, and stretching were of interest to us as the most common ones. Rather than focusing on the meaning of each action or the reasoning behind it, we brought our attention to the sounds that their voice emanated and its spatial dimension (e.g.~contagious laughter that spreads across the room, vocal warm-ups done in duets or triplets, loud abrupt breaths during breathing exercises). We then recreated those sounds using the 1st author's voice, where the spatial dimension was achieved with both the physical layout of the virtual singers (large circle), and with spatial sound panning methods. 

While this rendition of the work has not yet been shown to the public, it underwent several testing rounds where people could experience the installation, interact with the system (Fig.~\ref{fig:interaction}) and provide informal feedback. Almost all provided oral feedback mentioned a strong sense of the artist's presence in the room, so as the feeling of intimacy with some of the virtual singers. The combination of the ability to physically interact and provoke a response in each one of the singers, as well as the act of attentive listening, viewers felt engaged in a reciprocal, and genuine interaction.       

\section{Conclusion}
In this paper we have presented \emph{Transhuman Ansambl}, a novel interactive work that features the use of the human voice as the primary means of communication between humans and machines, using vocal articulations and song as the medium of interaction and creative exchange of human and machine agency \cite{bown_creative_2011}. Rather than conceptualising voice interaction in terms of commands or linguistic exchange, the \emph{Ansambl} requires a far more creative and open-ended approach to the interface. Developed primarily as a live performance, although some modes of the system resemble ``live looping'', what differentiates the two are distinct features such as i) partial control, ii) spatial dimension, and iii) immersion. To perform effectively with this system demands something of the performer as well as the virtual singers while both hold a certain level of autonomy. When singing with the \emph{ansambl}, the human performer can only decide on what to sing, while each virtual singer decides on the playback section. Consequently, the loss of absolute control over the playback creates space for improvisation and surprise. Additionally, the spatial dimension of the work creates an intimate space even when placed within a public live performance space. The act of surrounding the audience with all 16, presence-sensitive virtual singers and thus closer to the human performer, suggests a new way of employing technology to challenge the roles of performer-listener, and of ways we affect each other through sound and physical presence. The standalone sound installation mode of the work that allows the audience to interact with the virtual singers without the presence of the human performer is currently in the final stages of development.

\bibliographystyle{abbrv}

	\bibliography{Lucija.bib}

%
%




\end{document}